\documentclass[aps,twocolumn,superscriptaddress,citeautoscript,floatfix,showpacs]{revtex4-1}

\pdfoutput=1
\usepackage[english]{babel}
\usepackage[utf8]{inputenc}
\usepackage{amsmath}
\usepackage{graphicx}
\usepackage[dvipsnames]{xcolor}
\usepackage{amssymb}
\usepackage{subfigure}
\usepackage{anysize}
\usepackage{physics}
\usepackage{makeidx}
\usepackage{todonotes}
\usepackage{float}
\usepackage{xcolor,colortbl}
\usepackage{listings}
\usepackage{enumitem}
\usepackage[group-separator={,}]{siunitx}

\usepackage[paperwidth=210mm,paperheight=297mm,centering,hmargin=2cm,vmargin=2.5cm]{geometry}

\graphicspath{{images/}}

\definecolor{darkblue}{rgb}{0.1,0.2,0.6}
\definecolor{darkred}{rgb}{0.8,0.1,0.2}
\usepackage[colorlinks,citecolor=darkblue,linkcolor=darkred,urlcolor=darkblue]{hyperref}	
\usepackage[all]{hypcap} 

\providecommand{\newoperator}[2]{\newcommand*{#1}{\mathop{\mathrm{#2}}\nolimits}}
\newoperator{\sgn}{sgn}
\newoperator{\arctanh}{arctanh}
\newoperator{\argmax}{argmax}
\newoperator{\diag}{diag}
\newoperator{\spn}{span}

\def\ket#1{|#1\rangle}
\def\braket#1#2{\langle#1|#2\rangle}

\begin{document}

\title{Eigenstates hybridize on  all length scales at the many-body localization transition}

\def\urbana{
	Institute for Condensed Matter Theory and IQUIST and Department of Physics, 
	University of Illinois at Urbana-Champaign, Urbana, IL 61801, USA
}

\author{Benjamin Villalonga}
	\email{vlllngc2@illinois.edu}
	\affiliation{\urbana}
    
\author{Bryan K. Clark}
	\email{bkclark@illinois.edu}
	\affiliation{\urbana}

\date{\today}

\begin{abstract}
An interacting quantum system can transition from an ergodic to a many-body localized (MBL) phase under the presence of sufficiently large disorder.
Both phases are radically different in their dynamical properties, which are characterized by highly excited eigenstates of the Hamiltonian.
Each eigenstate can be characterized by the set of quantum numbers over the set of (local, in the MBL phase) integrals of motion of the system.  
In this work we study the evolution of the eigenstates of the disordered Heisenberg model as the disorder strength, $W$, is varied adiabatically.
We focus on the probability that two `colliding' eigenstates  hybridize as a function of both the range $R$ at which they differ as well as the strength of their hybridization.
We find, in the MBL phase, that the probability of a colliding eigenstate hybridizing strongly at range $R$ decays as $Pr(R)\propto \exp [-R/\eta]$, with a length scale $\eta(W) = 1 / (B \log(W / W_c) )$ which diverges at the critical disorder strength $W_c$.
This leads to range-invariance at the transition,
suggesting the formation of resonating cat states at all ranges.
This range invariance does not survive to the ergodic phase, where hybridization is exponentially more likely at large range, a fact that can be understood with simple combinatorial arguments.
In fact, compensating for these combinatorial effects allows us to define an additional correlation length $\xi$ in the MBL phase which is in excellent agreement with previous works and which takes the critical value $1 / \log(2)$ at the transition, found in previous works to destabilize the MBL phase.
Finally, we show that deep in the MBL phase hybridization is dominated by two-level collisions of eigenstates close in energy. 
\end{abstract}


\maketitle

\textbf{Introduction.---}
Certain disordered interacting systems can transition to the many-body localized (MBL) phase, breaking ergodicity and defying thermalization in the process~\cite{fleishman_interactions_1980,basko_metalinsulator_2006,gornyi_interacting_2005,rigol_thermalization_2008,nandkishore_many-body_2015,luitz_many-body_2015,luitz_long_2016}.
A fully MBL Hamiltonian, like the sufficiently disordered one-dimensional Heisenberg model (Eq.~\eqref{eq:model}), can be diagonalized by a unitary operator $U$ which can be expressed as a low-depth quantum circuit~\cite{serbyn_local_2013,huse_phenomenology_2014,pollmann_efficient_2016,pekker_encoding_2017}.
This results in localized eigenstates with area law entanglement~\cite{bauer_area_2013,yu_bimodal_2016,kjall_many-body_2014}, \emph{i.e.}, they can be disentangled to product states through a small number of local transformations.
By continually decreasing the strength of the disorder, the system's eigenstates evolve transitioning into ergodic eigenstates at a sharp value of the disorder strength; these eigenstates are delocalized and have volume law entanglement.
Furthermore, while MBL eigenstates do not experience level repulsion and follow Poisson statistics, ergodic eigenstates follow GOE~\cite{oganesyan_localization_2007,luitz_many-body_2015,buijsman_random_2019}.
For recent reviews on MBL see Refs.~\cite{nandkishore_many-body_2015,altman_universal_2015,luitz_ergodic_2017,imbrie_local_2017,agarwal_rare-region_2017,abanin_recent_2017,alet_many-body_2018}.

For a spin-$1 \over 2$ system whose Hamiltonian $H$ is diagonalized by the unitary matrix $U$, the two-level Hermitian operators
\begin{equation}
\label{eq:tau}
    \tau^z_i \equiv U \sigma^z_i U^\dagger
\end{equation}
commute with $H$ and with each other.
In MBL, these operators (often called $\ell$-bits) have local support and can be interpreted as locally dressed spins.
$\ell$-bits have been a key quantity in understanding the phenomenology of the MBL phase~\cite{serbyn_local_2013,huse_phenomenology_2014,pollmann_efficient_2016,pekker_encoding_2017}, and considerable effort has been devoted to numerically access $\ell$-bits~\cite{pekker_encoding_2017,pekker_encoding_2017,inglis_accessing_2016,rademaker_many-body_2017,kulshreshtha_behavior_2018,wahl_efficient_2017,wahl_signatures_2019,thomson_time_2018,pancotti_almost_2018}.
Using $\ell$-bits we can identify each eigenstate $\ket{n}$ of $H$ via the $L$ pseudo-spins corresponding to the eigenvalues of the $L$ $\ell$-bits $\{ \tau^i_z \}$.
Alternatively, these eigenvalues correspond to the spin configuration of the product state generated by $U^\dagger \ket{n}$. 

\begin{figure*}[t]
\centering
\includegraphics[width=2.00\columnwidth]{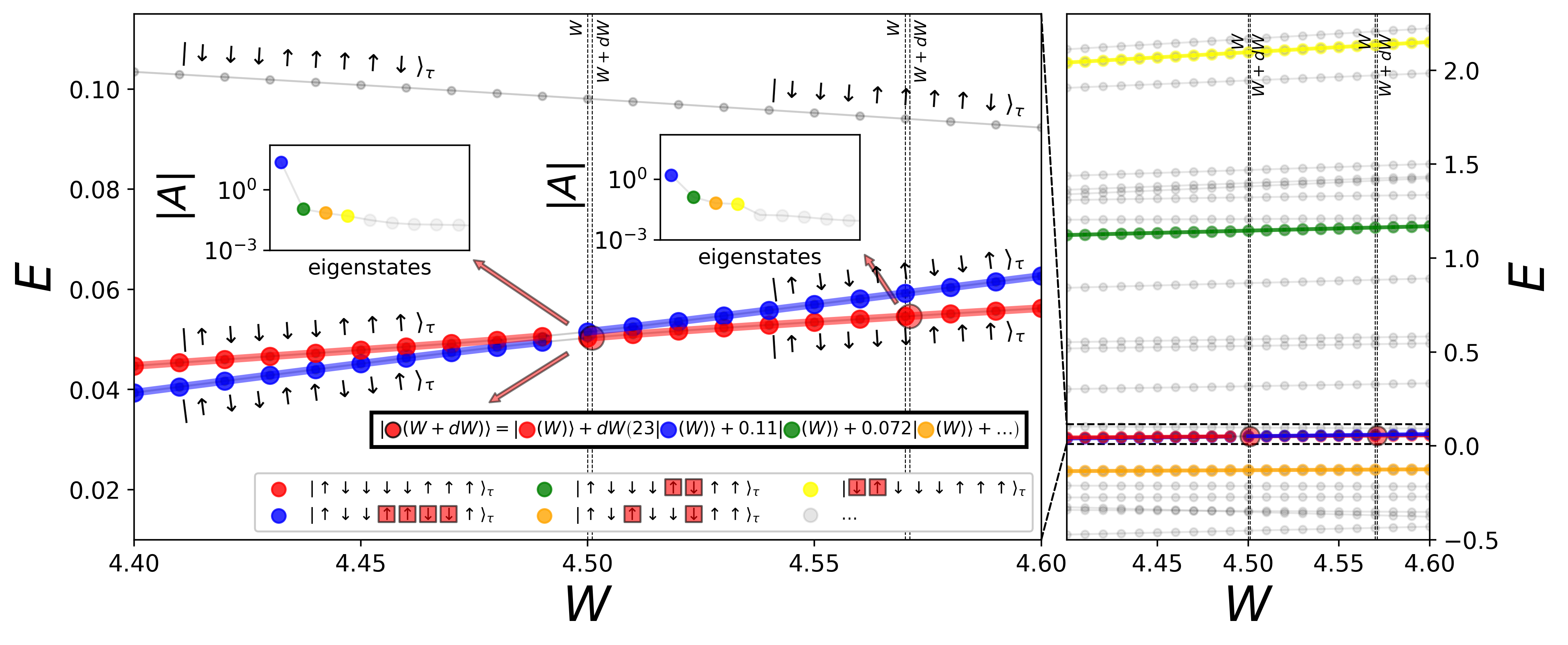}
\caption{\label{fig:figure0}
Example of colliding eigenstates in the interval $4.4<W<4.6$ for $L=8$.
Dotted lines show the values of $W$ where the hybridization is computed, as well as $W+dW$, for $dW = 10^{-7}$ (exaggerated in distance in the figure to make it visible).
Colors indicate different quantum number configurations.
The insets show the hybridization strength $|A|$ between the red and \emph{other} eigenstates whose range from the red eigenstate is 1 (green and yellow; see pseudo-spins highlighted in the legend) and 3 (orange and blue); note that the amplitudes of the adiabatically evolved wave function are obtained by multiplying these values by $dW$.
Notice this differs from the Hamming distance.  
Notice also that strongly coupled eigenstates range 1 away from the red eigenstate are $\mathcal{O}(1)$ distant in energy, since they are easily coupled through the term $\sigma_i^z \sigma_{i + 1}^z$ in Eq.~\eqref{eq:model} through a single (pseudo-)spin flip.
On the contrary, eigenstates a larger range away hybridize only when close in energy to the red eigenstate.
} 
\end{figure*}

We focus on the one-dimensional spin-$1 \over 2$ nearest-neighbor antiferromagnetic Heisenberg chain with random on-site magnetic fields:
\begin{align}
\label{eq:model}
	H(W) = \frac{1}{4} \sum_{i=0}^{L-2} \vec{\sigma}_i \cdot \vec{\sigma}_{i+1} - \frac{W}{2} \sum_{i=0}^{L-1} h_i \sigma^z_i \text{.}
\end{align}
The on-site magnetic fields $\left\{ h_i\right\}$ are sampled uniformly at random from $\left[ -1, 1 \right]$ and $W$ is the disorder strength.
The model of Eq.~\eqref{eq:model} has been studied extensively in the context of MBL~\cite{oganesyan_localization_2007,znidaric_many-body_2008,berkelbach_conductivity_2010,pal_many-body_2010,bauer_area_2013,luitz_many-body_2015,bar_lev_absence_2015,agarwal_anomalous_2015,bera_many-body_2015,luitz_extended_2016,luitz_long_2016,luitz_anomalous_2016,luitz_information_2017,yu_bimodal_2016,khemani_critical_2017,khemani_two_2017,serbyn_power-law_2016,de_tomasi_quantum_2017,herviou_multiscale_2019,gray_many-body_2018,gray_scale_2019,laflorencie_chain_2020}.
It presents an ergodic-MBL transition at infinite temperature (middle of the energy spectrum) at a critical $W_c \approx 3.8$~\cite{luitz_many-body_2015}.  
Numerical simulations also suggest a transition $W$ that varies with energy, forming a mobility edge~\cite{luitz_many-body_2015,villalonga_exploring_2018}.
This Hamiltonian conserves total magnetization $\sum_i \sigma^z_i$; throughout this work we focus on the zero magnetization sector.

While many aspects of the MBL phase are well understood, much less is known about the transition, where locality must break down.
Renormalization group studies suggest the proliferation of ergodic grains or resonating clusters in the system near the transition~\cite{vosk_theory_2015,potter_universal_2015,dumitrescu_kosterlitz-thouless_2019}; furthermore, Refs.~\cite{de_roeck_stability_2017,luitz_how_2017} suggest that the inclusion of ergodic grains destabilizes the MBL phase when their length exceeds the critical correlation length $\xi_c = 1 / \log(2)$. 
In addition, based on the presence of ergodic grains close to the transition, the theory of  Refs.~\cite{thiery_microscopically_2017,thiery_many-body_2018} predicts the emergence of a length scale that diverges at the transition.

\begin{figure}[t]
\centering
\includegraphics[width=1.00\columnwidth]{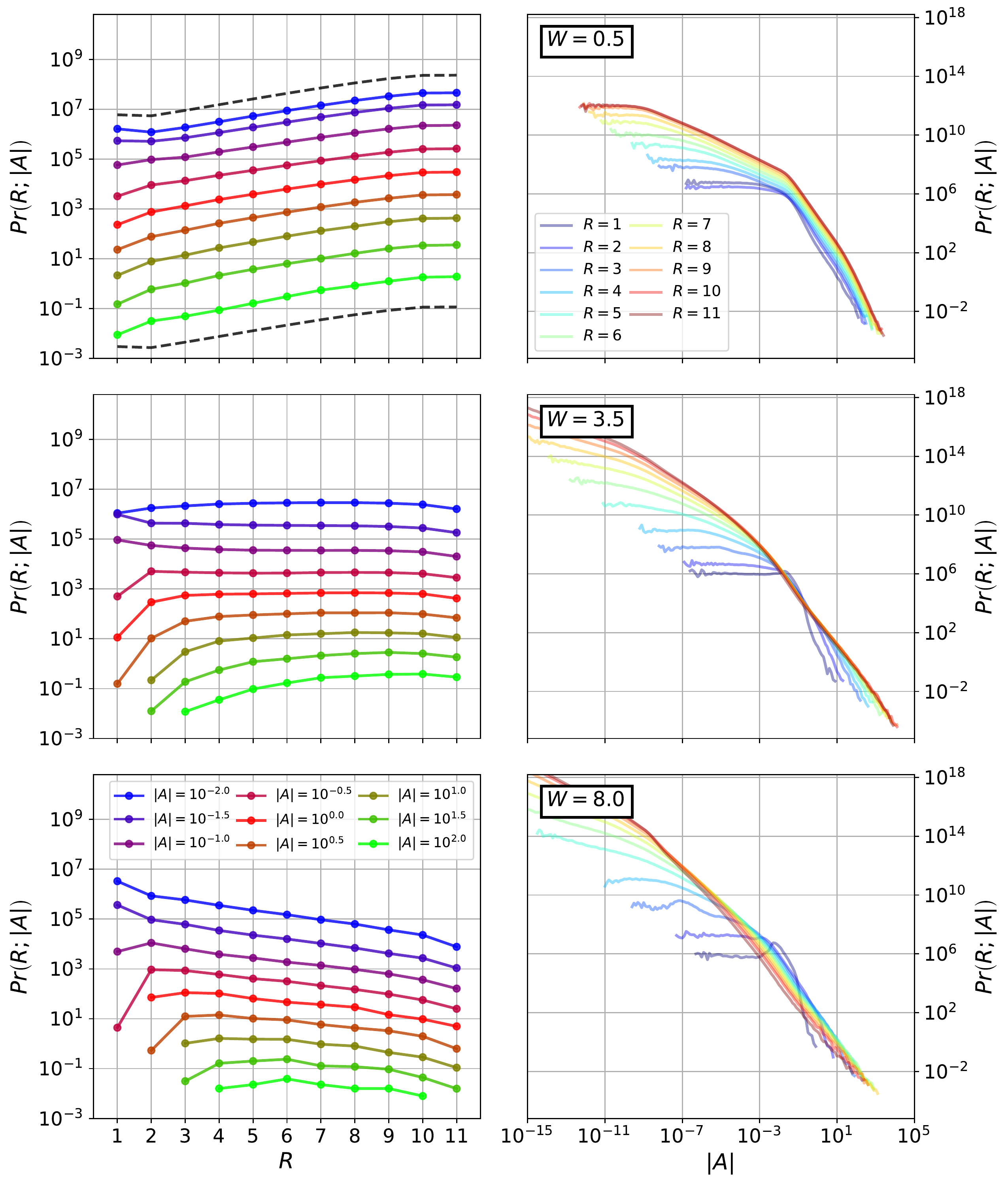}
\caption{\label{fig:invariance} Probability $Pr(R, |A|)$ that two eigenstates hybridize over any cluster of range $R$ with hybridization rate $|A|$, for all pairs of eigenstates over 1024 different disorder realizations on a chain of size $L=12$.
We can observe three distinct behaviors.
In the ergodic phase ($W = 0.5$) contributions from long ranges are dominant due to a simple combinatorial argument.
At the transition ($W  = 3.5$) all ranges contribute equally, and distributions collapse to a universal curve at large enough strength and range.
In the MBL phase ($W = 8.0$), short range exponentially dominates, in line with the concept of localization.} 
\end{figure}

In this work, we consider the evolution of the eigenstates of $H(W)$ as we tune the disorder strength $W$ from the MBL phase at large disorder, to the transition, and into the ergodic phase at small disorder.
During this evolution, driven by `collisions', each eigenstate hybridizes onto the other eigenstates of the system (see Fig.~\ref{fig:figure0} for an example), changing their properties and strongly influencing the distinct level statistics of the MBL and ergodic phases.
This hybridization occurs over a set of spins spanning a certain size or range $R$.
For strongly hybridized eigenstates there is universal behavior for these collisions.  
The key results of this work are as follows. 
\begin{itemize}
    \item At the transition, we find the probability that two eigenstates hybridize over a set of spins is \emph{range invariant} (Fig.~\ref{fig:invariance} (middle-left)) losing locality in a qualitatively different way from the ergodic phase (where hybridization over all clusters of spins is equally likely); this suggests the formation of singlets at the transition at all length scales.
    \item In the MBL phase, the probability of hybridization decays exponentially with range and we identify a diverging length scale $|\alpha(W)|^{-1}$ (see Eq.~\eqref{eq:alpha}) from this exponential decay.
    Additionally, hybridization deep in the MBL phase is dominated by sporadic pair-wise collisions.
    \item We identify, and specify the functional form of, a non-diverging correlation length $\xi(W)$ (Eq.~\eqref{eq:xi}), which takes the expected value $\xi_c = 1 / \log(2)$~\cite{de_roeck_stability_2017,luitz_how_2017} at the transition and closely agrees, throughout the MBL phase, with correlation lengths computed in other capacities~\cite{villalonga_exploring_2018}.
\end{itemize}

\textbf{Adiabatic evolution of $\ket{n(W)}$ and range $R$ hybridization.---}
Given a fixed disorder realization, \emph{i.e.}, a set of randomly sampled magnetic fields $\left\{ h_i \right\}$, the model of Eq.~\eqref{eq:model} defines a Hamiltonian $H(W)$ with disorder strength $W$.
When varying $W$ adiabatically (note that this is different from the local, adiabatic perturbation of Ref.~\cite{khemani_nonlocal_2015}), the eigenstates of $H(W)$ evolve continuously and we can label them by their position in the energy spectrum.
Alternatively, we can label these eigenstates by their set of quantum numbers, $\ket{n} = \ket{\uparrow \downarrow \downarrow \ldots}_\tau$, where the subscript $\tau$ indicates that this is a product state in the diagonal basis of $\ell$-bits.
In this work, we make use of this labelling by quantum numbers.

Let us start by defining hybridization and the concept of \emph{hybridization rate} (see Fig.~\ref{fig:figure0} for an example).
Given a differential change $W \rightarrow  W + dW$, we induce the trajectory $\ket{n(W)} \rightarrow \ket{n(W + dW)}$, in which the eigenstate slightly rotates in Hilbert space.
In the basis of eigenstates $\left\{ \ket{k(W)} \right\}$ of $H(W)$ we can now write 
\begin{equation}
     \ket{n(W + dW)} = c_n^n \ket{n(W)} + \sum_{k\neq n}^{\text{dim}(H)} A_k^{n}dW |k(W)\rangle
\end{equation}
where, by normalization, $c_n^n = 1-\sqrt{\sum_{k\neq n}^{\textrm{dim} }(A_k^n dW)^2}$.
We call $A^n_k$, with $k \neq n$, the \emph{hybridization rate} of $\ket{n(W)}$ with $\ket{k(W)}$. 
Throughout this paper, we use $dW=10^{-7}$.
Note that $\left( A^n_k dW \right)^2$ can also be thought of as a transition probability from $\ket{n(W)}$ to $\ket{k(W)}$.
We will be concerned with the probability of hybridization at various strengths of $|A|$ (note we have dropped indices $n$ and $k$ for simplicity) and ranges $R$, $Pr(R; |A|)$ (see Appendix~\ref{sec:obtaining_pr} for details on the computation of this distribution numerically).

We now define the spatial range $R$ associated with a pair of eigenstates $\ket{n(W)}$ and $\ket{k(W)}$.
To do so, we look at the difference in their quantum numbers, \emph{i.e.} their eigenvalues with respect to the $\ell$-bit operators $\left\{ \tau_i^z \right\}$.
These $\ell$-bits are determined via the diagonalizing unitary $U$ by Eq.~\ref{eq:tau}.
Note that there are many unitaries which diagonalize $H$ which differ by permuting the columns of $U$, as well as changing their signs.
Making the right choice of $U$ is essential to construct the most localized possible set of $\ell$-bits.
While finding the optimal $U$ is not efficiently solvable (and indeed depends on metric of localization chosen), the Wegner-Wilson flow approach described in Ref.~\cite{pekker_fixed_2017} gives good results in practice.
In the present work, we always refer to the set of operators $\{\tau^z_i\}$ and to the unitary $U$ found by this method.

We say that two eigenstates of a spin Hamiltonian $H$, $\ket{n}$ and $\ket{k}$, have a difference of range $R$ if the left-most and the right-most disagreeing quantum numbers between the two are $R$ pseudo-sites away from each other.
For example, eigenstates $\ket{n(W)} = \ket{\uparrow \downarrow \downarrow \uparrow \downarrow \uparrow}_\tau$ and $\ket{k(W)} = \ket{\uparrow \uparrow \uparrow \downarrow \downarrow \downarrow}_\tau$ differ on a cluster of range $R = 4$ (between the second and the sixth pseudo-spins).
While pseudo-spins are not exactly spins on real space, they are locally dressed spins in the MBL phase, and so the notions of distance in real space and in the space induced by l-bits are related.
\\

\textbf{Results.---}
When two eigenstates collide, their disagreeing pseudo-spins  hybridize, while all other pseudo-spins which are common to both states remain approximately frozen in their configuration.
This hybridization drives the eigesntates towards the formation of a cat state across a region of range $R$.
We are interested in two related probabilities: $Pr(R, |A|)$ is the joint probability that a pair of eigenstates $\ket{n(W)}$ and $\ket{k(W)}$ hybridize over \emph{any} cluster of range $R$ with a hybridization rate of magnitude $|A^n_k|$.  $C(R;|A|)$ is the probability that a \emph{particular cluster} of range $R$ hybridizes with magnitude $|A|$.
Note that in the thermodynamic limit
\begin{equation}
\label{eq:C}
    C(R;|A|) = Pr(R;|A|)/2^R \text{,}
\end{equation}
\emph{i.e.}, the probability that a particular range-$R$ cluster hybridizes goes as the probability \emph{any} range-$R$ cluster hybridizes over the number of clusters with range $R$ (see Appendix~\ref{sec:combinatorial}).

\begin{figure}[t]
\centering
\includegraphics[width=1.00\columnwidth]{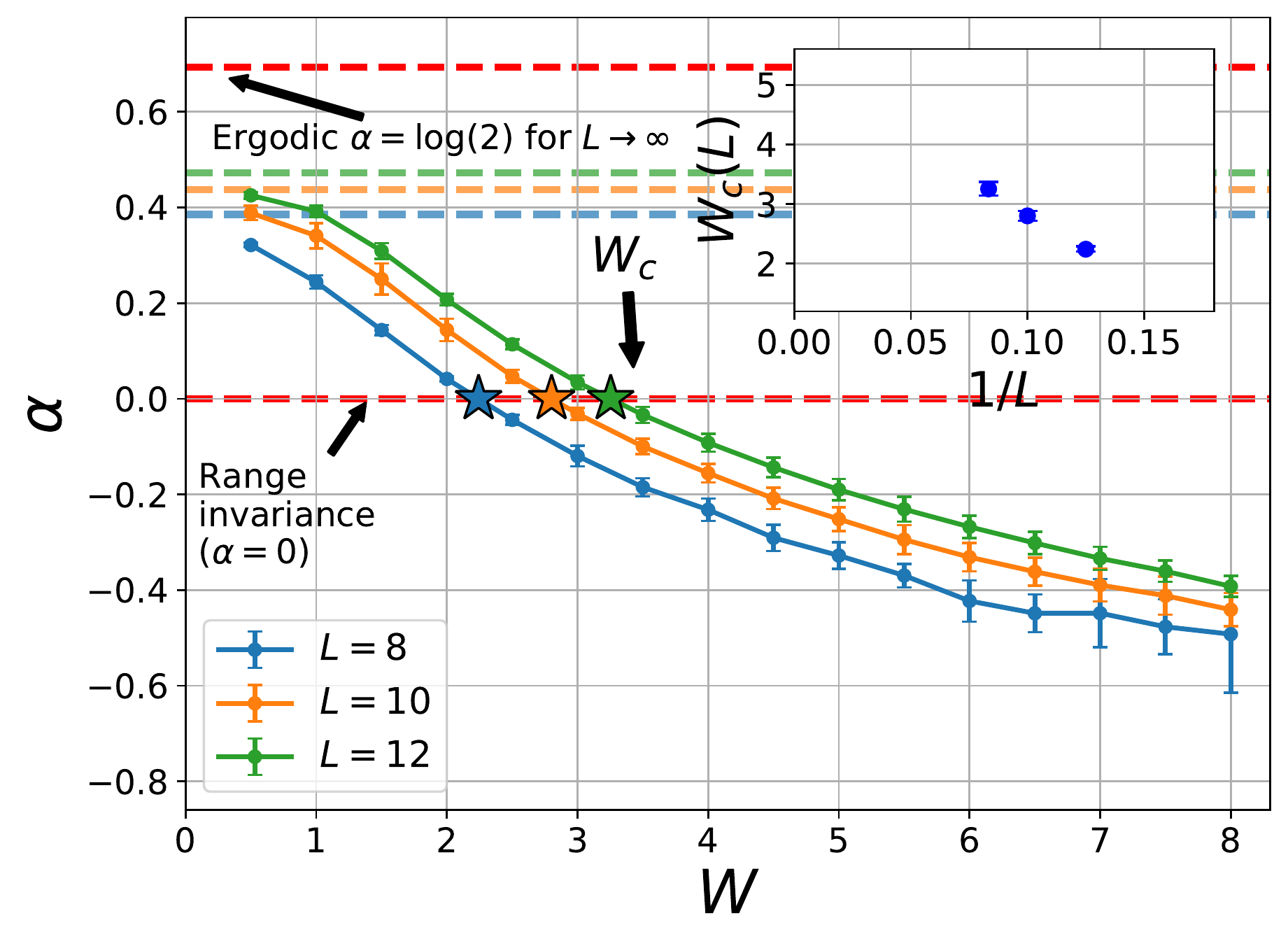}
\includegraphics[width=1.00\columnwidth]{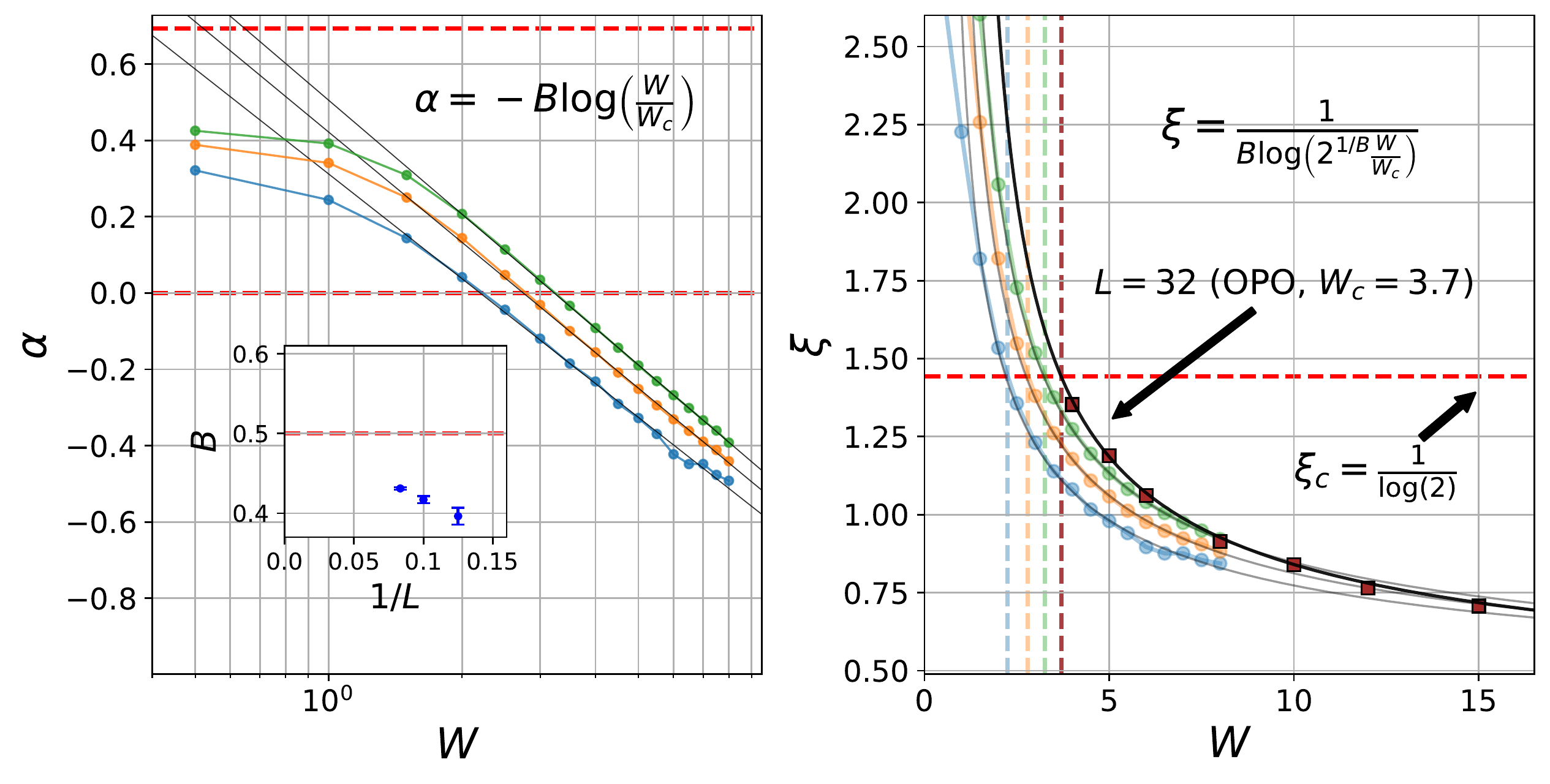}
\caption{\label{fig:alpha}
\textbf{Top:} Exponent $\alpha(W)$ of the scaling of $Pr(R; |A|) \propto \exp(\alpha R)$ at constant $|A|$ (see left panel of Fig.~\ref{fig:invariance}) for systems of size $L=8, 10, 12$ computed by averaging the slope obtained from a linear fit of $\log(Pr(R; |A|))$ as a function of $R$, for all values of $|A|$ in the interval $[10^{-1.5}, 10^{-0.5}]$ ($[10^{-1}, 10^{-0.5}]$ for $L = 8$); errorbars represent their standard deviation.
Critical points $W_c(L)$ (stars and inset) defined by $\alpha(W_c) = 0$ (range invariance in $Pr(R, |A|)$).
Dashed lines show the expected $\alpha$ if the probability of hybridization between any two spin configurations were constant (see Appendix~\ref{sec:combinatorial}).
This goes to $\log(2)$ (top red line) in the thermodynamic limit.
\textbf{Bottom left:} Fit of the form $\alpha = -B \log[W/W_c(L)]$ for $W > W_c(L)$ with inset showing values of $B$ as a function of $1/L$; red-dashed line is $B=1/2$.
The fit breaks down deep in the ergodic phase.
\textbf{Bottom right:} Black lines show the localization length $\xi(W)$ of Eq.~\eqref{eq:xi} for $L = 8, 10, 12, 32$; for $L = 32$ we use $W_c = 3.7$ and $B = 1 / 2$.
Brown squares show the correlation length of one-particle orbitals (OPO) as reported in Ref.~\cite{villalonga_exploring_2018} (Fig.~4, $L = 32$, energy density $\epsilon=0.5$); circles show $\xi(W) = [-\alpha(W)^{-1} + \log(2)]^{-1}$ for $L = 8, 10, 12$.
Vertical dashed lines show the value of $W_c$ for each system size; $\xi = 1 / \log(2)$ at $W = W_c$.
\textbf{All panels:} We have tested averaging various different spans of $R$ and $|A|$.
The logarithmic dependence of $\alpha$ for $W \geq W_c$ appears very robust to all choices; the particular values of $W_c$ and $\beta$ for a given $L$ can change at the $20\%$ level depending on the choice of fit.}
\end{figure}

Fig.~\ref{fig:invariance} (left column) shows slices of $Pr(R, |A|)$ as a function of range $R$ at constant (strong) hybridization rates $|A|$.
It can be seen that this probability scales exponentially with range $R$ for all $W$ as 
\begin{equation}
    Pr(R; |A|) \propto \exp\left[ \alpha(W) R \right]
\end{equation} 
The slope of the semi-log plots in Fig.~\ref{fig:invariance} (left), $\alpha(W)$, is shown in Fig.~\ref{fig:alpha} (top).

In the ergodic phase, $\alpha > 0$; furthermore, deep in the ergodic phase $Pr(R; |A|)$ is proportional to the number of clusters of range $R$ (see dashed lines of Fig.~\ref{fig:alpha} (top)), which scales as $2^R$ in the thermodynamic limit, and slower than that in finite systems (see Appendix~\ref{sec:combinatorial}).  This suggests the notion of locality is completely lost in the ergodic phase, and the probability, $C(R; |A|)$,  that \emph{a particular cluster} of pseudo-spins hybridizes is constant, independent of its range $R$.

In the MBL phase and around the transition $\alpha(W)$ scales as 
\begin{equation} 
\label{eq:alpha}
    \alpha(W) = -B \log(W/W_c) \text{,}
\end{equation}
as seen in Fig.~\ref{fig:alpha} (bottom left).
$B$ is a constant (shown in the inset) and numerically trends toward $1/2$.
$W_c$ is the critical point identified by where $\alpha(W_c) = 0$; note the logarithmic fit breaks deep in the ergodic phase.
The critical point $W_c$ trends upward as a function of $L$, reaching a thermodynamic value above $W\geq 3.25$.
As expected for a localized state, $\alpha < 0$ in the MBL phase, and the probability that \emph{any cluster} ($Pr(R; |A|)$) of range $R$ hybridizes is exponentially suppressed with $R$.
Notice that, in the MBL phase, this gives an algebraic dependence with $W$ which goes as $Pr(R, |A|) \propto (W/W_c)^{-BR}$.

Strikingly, at the transition $Pr(R, |A|)$ becomes range-invariant at strong values of $|A|$ out to a range-dependent cutoff in $|A|$ after which $Pr(R, |A|)$ drops the universal, range-independent behavior (see right column of Fig.~\ref{fig:invariance}); the larger $R$, the larger the cutoff.
Notice this is consistent with Eq.~\eqref{eq:alpha} at $W = W_c$ ($\alpha(W_c) = 0$) and allows us to identify a length scale $\eta(W) = [-\alpha(W)]^{-1}$ for $W \geq W_c$ that diverges as $W \rightarrow W_c$.
It should be emphasized that this is a special type of `cluster-agnostic range-invariance' which differs qualitatively from the sort of range-invariance found in the ergodic phase.
At the transition, $Pr(R;|A|)$ is constant, while in the ergodic phase $C(R;|A|)$ is constant. 

It is also possible to define a correlation length $\xi(W)$ from $C(R; |A|) = \exp[-R/\xi(W)]$ at large, fixed $|A|$.
Working in the thermodynamic limit, we have that 
\begin{align} 
\label{eq:xi}
    \xi(W)^{-1} &=  \eta(W)^{-1} + \xi_c^{-1}  \nonumber \\
                &=  \left[B \log (2^{1/B} \frac{W}{W_c} )  \right]^{-1}
\end{align}
where the second equation follows from our numerically determined functional form for $\alpha(W)$. 
The critical correlation length $\xi_c = \xi(W_c) = \log(2)^{-1}$ is in agreement with the value that Refs.~\cite{de_roeck_stability_2017,luitz_how_2017} find to destabilize the MBL phase
Note that in our case the value of $\xi_c$ comes from positing `cluster-agnostic range-invariance' at the critical point (where $C(R; |A|)$ becomes independent of $R$), and is independent of the functional form of $\alpha(W)$ and requires no additional assumptions about ergodic grains.

Using $B = 1/2$ and a single fitting parameter, $W_c = 3.7$, we compare $\xi(W)$ of Eq.~\eqref{eq:xi} with the correlation lengths found in Ref.~\cite{villalonga_exploring_2018}, which were computed from the decay of approximate $\ell$-bits generated from $L=32$ MBL eigenstates.
The comparison is shown in Fig.~\ref{fig:alpha} (bottom right) and the agreement is striking.  
\\

\textbf{Pair-wise collisions in MBL.---}
We now show that, in the MBL phase, the probability of colliding (at fixed $R$) with a given strength $|A|$ (hybridization rate) decays as a power law with strength $Pr(|A|;R) \propto |A|^{-3/2}$.
This algebraic decay comes from the avoided level crossings of two eigenstates, suggesting that hybridization in the MBL phase is dominated by sporadic pair-wise collisions (note this is related to the resonating pairs of configurations found in Ref.~\cite{gopalakrishnan_low-frequency_2015} to be responsible for the conductivity deep in the MBL phase as well as the rare resonances considered of Ref.~\cite{kjall_many-body_2018}). 
On the contrary, ergodic eigenstates are continuously hybridizing through overlapping processes, which are due to
continuous collisions.

\begin{figure}[t]
\centering
\includegraphics[width=1.00\columnwidth]{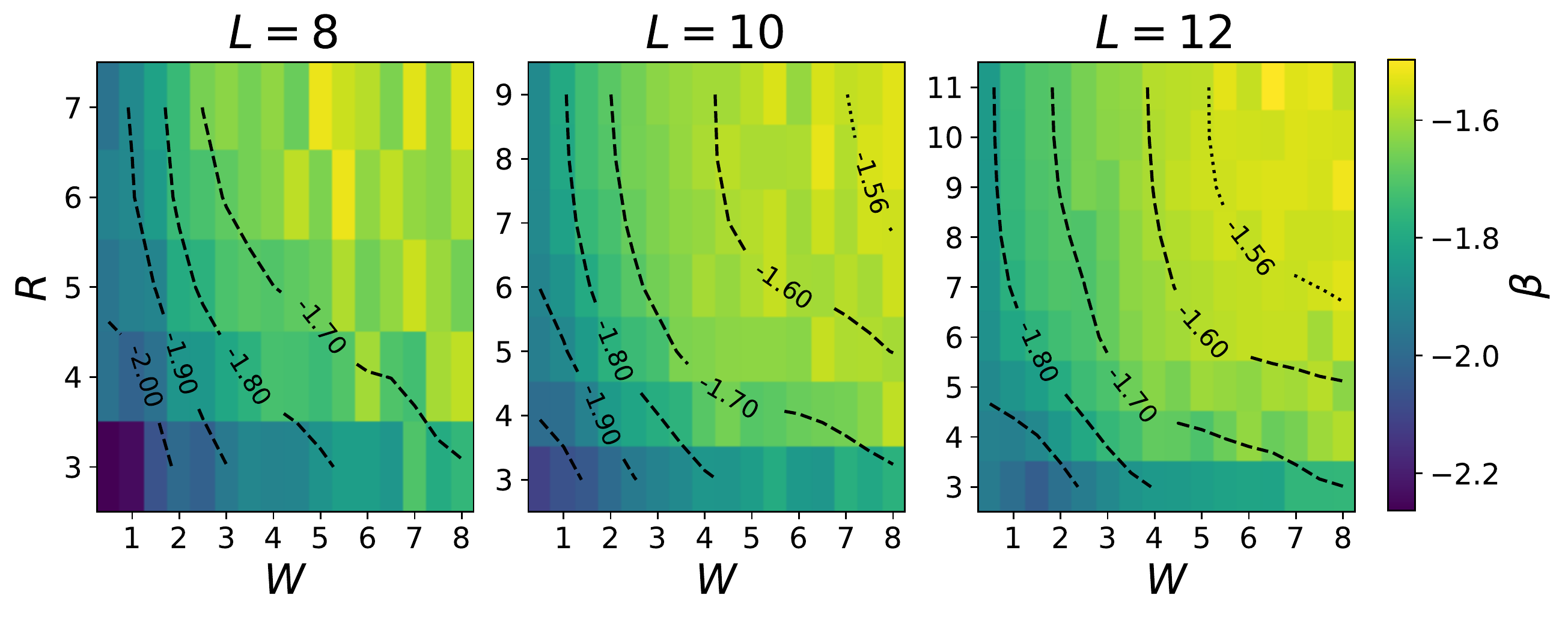}
\includegraphics[width=1.00\columnwidth]{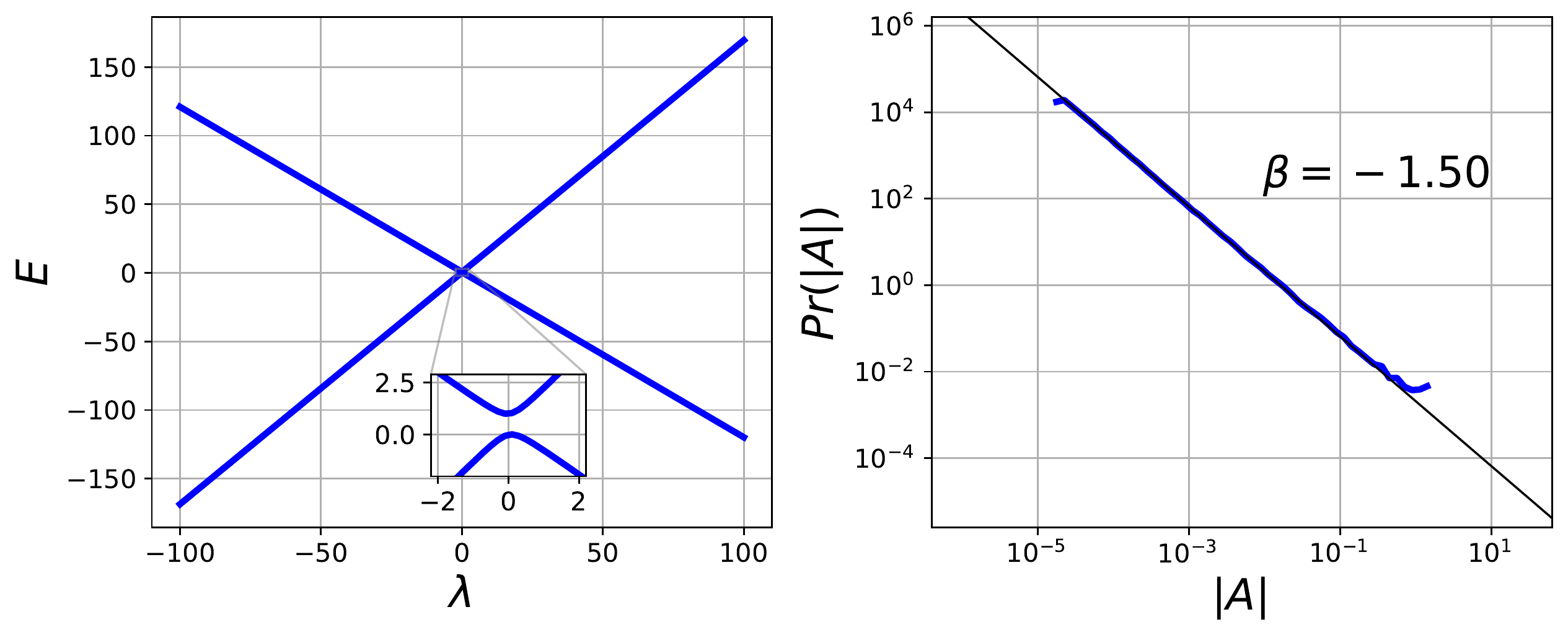}
\caption{\label{fig:beta} \textbf{Top:}
colormap with the exponent $\beta$ of the power law decay of $Pr(R, |A|) \propto |A|^\beta$ at strong $|A|$ for different values of $R$ and $W$.
We can see that $\beta \rightarrow -{3 \over 2}$ at large $R$ and strong $W$.
\textbf{Bottom:} level collision of the two-level toy model $H_\text{toy}(\lambda) = H_0 + \lambda H_1$ and its corresponding $Pr(|A|)$.
We recover the exponent $\beta = -{3 \over 2}$, which suggests that strong, long-range hybridization in MBL is dominated by rare, pair-wise collisions.
} 
\end{figure}

We compute the exponent $\beta$ of the universal power law decay of $Pr(R, |A|) \propto |A|^\beta$, for constant (large) $R$, right before $Pr(R, |A|)$ hits the cutoff characteristic of each range $R$ (see Fig.~\ref{fig:invariance} (right) and Appendix~\ref{sec:beta_fits}).

The top panel of Fig.~\ref{fig:beta} shows the values of $\beta$ as a function of $W$ and $R$ for different system sizes $L$.
At large $W$ and $R$, we find $\beta \rightarrow -{3 \over 2}$.
The exponent $\beta = -{3 \over 2}$ can be obtained from a toy two-level model $H_\text{toy}(\lambda) \equiv H_0 + \lambda H_1$ for which two states $\ket{1(\lambda)}$ and $\ket{2(\lambda)}$ undergo a collision.
As can be seen in the bottom panel of Fig.~\ref{fig:beta}, by sampling hybridization rates between $\ket{1}$ and $\ket{2}$, $|A| = |\braket{1(\lambda)}{2(\lambda + d\lambda)}|$ uniformly at random over a large window in $\lambda$ ($[-\lambda_m, \lambda_m\equiv 100]$), we precisely obtain a probability distribution $Pr(|A|)$ that decays as a power law with exponent $\beta = -{3 \over 2}$.
This behavior mimics the rare pair-wise collisions of the MBL phase.
\\

\textbf{Conclusions.---}
In this work, we have considered the hybridization of eigenstates as they are driven from the MBL phase, through the transition, and into the ergodic phase.
In particular, we have studied two related probabilities: $Pr(R; |A|)$ and $C(R; |A|)$.
$Pr(R; |A|)$ is the probability that a pair of eigenstates hybridize over a region of range $R$ (for fixed and strong hybridization rate or strength $|A|$), while $C(R; |A|)$ is the probability that they hybridize over a specific cluster of pseudo-spins spanning a region of range $R$.
While both probabilities are suppressed exponentially with $R$ in the MBL phase, $Pr(R; |A|)$ is range invariant at the transition, and $C(R; |A|)$ is range invariant deep in the ergodic phase.
This lets us define two length scales: $\eta(W)$ and $\xi(W)$.
$\eta(W)$ is related to $Pr(R; |A|)$ and represents a length scale that diverges at the transition.
$\xi(W)$ is related to $C(R; |A|)$; it has the anticipated correlation length $1/\log 2$ at the transition and is a surprisingly close match to other correlation lengths computed in the MBL phase. 
We identify a functional form for $Pr(R; |A|)$ in the MBL phase and around the transition, which can be extended to functional forms for $C(R; |A|)$, $\eta$ and $\xi$.
We anticipate that the ansatz of Eq.~\eqref{eq:xi} will be key in understanding the phenomenology of the MBL transitions.

The phenomenology of the transition is often thought of as being driven by the proliferation of resonances or cat states which are believed to be responsible for melting the MBL phase.
Our finding of range-invariant hybridization (at all large enough hybridization strengths) of eigenstates at the transition suggests the proliferation of singlets - cat states -  at all scales, which are related to the appearance of resonances in a range-invariant fashion across the system.
To identify the range-invariant hybridization and correlation lengths required us both to work in the pseudo-spin basis, abandoning strict locality in favor of a quasi-local view of the system, as well as separating out probability distributions by hybridization strengths so range-dependent cutoffs do not contaminate averages.
 
It is natural to expect locality to break down at the MBL-ergodic transition.
There are two reasonable ways this could happen: either $C(R; |A|)$ or $Pr(R; |A|)$ become range-invariant.
Our observations find that the transition corresponds to the latter, which alone is sufficient to set the value of $\xi_c = 1 / \log(2)$ found in the literature, without assumptions on the inclusion of ergodic grains.

Finally, Eq.~\ref{eq:xi} suggests the intriguing fact that the critical point $W_c$
could be determined by properties of the system deep in the MBL phase; it also suggests an additional way to interpret current correlation lengths in the literature.
We also note that $Pr(R;|A|)$ scales algebraically with $W$ and it is an interesting question whether this has any relationship to the algebraic scaling with $W$ seen in Ref.~\cite{yu_bulk_2019} for the smallest coupled energy scales within the MBL phase.

\begin{acknowledgments}
The authors would like to thank Vedika Khemani for pointing out the importance of clusters with more than two resonating sites as a signature of many-body effects.
We are grateful to David Luitz and Greg Hamilton for carefully reading and providing comments on our manuscript.
We are also thankful to Sarang Gopalakrishnan for pointing out the role of resonances in the conductivity in MBL, as well as providing a relevant reference.
We would like to acknowledge useful discussions with Greg Hamilton, Eli Chertkov, David Huse, Romain Vasseur, Chris Laumann, and Anushya Chandran.
We both acknowledge support from the Department of Energy grant DOE de-sc0020165.
BV also acknowledges support from the Google AI Quantum team.
This project is part of the Blue Waters sustained petascale computing project, which is supported by the National Science Foundation (awards OCI-0725070 and ACI-1238993) and the State of Illinois. Blue Waters is a joint effort of the University of Illinois at Urbana-Champaign and its National Center for Supercomputing Applications.
\end{acknowledgments}

\bibliographystyle{apsrev4-1}
\bibliography{main}

\clearpage
\appendix 

\section{Obtaining the joint probability distribution $Pr(R; |A|)$}
\label{sec:obtaining_pr}

In order to numerically obtain the joint probablility distribution $Pr(R; |A|)$ (see Fig.~\ref{fig:invariance} in the main text), we first compute the absolute value of all hybridization ratios between all pairs of eigenstates over 1024 random realizations of the Hamiltonian in Eq.~\eqref{eq:model} of the main text.
We do this for different disorder strengths $W$ and system sizes $L$.
Each hybridazation ratio is associated to a range $R \in [1, L - 1]$, and so for each tuple $(L, W, R)$ we now have a set of hybridization ratios (\emph{i.e.}, a set of values of $|A|$).

Then, for each $(L, W, R)$ we histogram the corresponding set with log-spaced bins (\emph{i.e.}, with bins that are equally spaced on a logarithmic scale); this gives us $h_{L, W, R}(|A|_i)$, where $|A|_i$ is a discrete variable that takes the middle point of bin $i$.
In order to precisely obtain a probability distribution, we now need to divide each $h_{L, W, R}(|A|_i)$ by the size of bin $i$; we denote the result by $f_{L, W, R}(|A|_i)$.
Finally, we enforce normalization for each tuple $(L, W)$ by dividing $f_{L, W, R}(|A|_i)$ by $S_{L, W} \equiv \sum_R \sum_{i} f_{L, W, R}(|A|_i)$, after which we obtain $Pr(R; |A|_i) \equiv f_{L, W, R} / S_{L, W}$, which for simplicity we just denote by $Pr(R; |A|)$.

\section{$Pr(R; |A|)$ for $L = 8, 10$}
\label{sec:pr_smaller_L}

\begin{figure}[t]
\includegraphics[width=1.0\columnwidth]{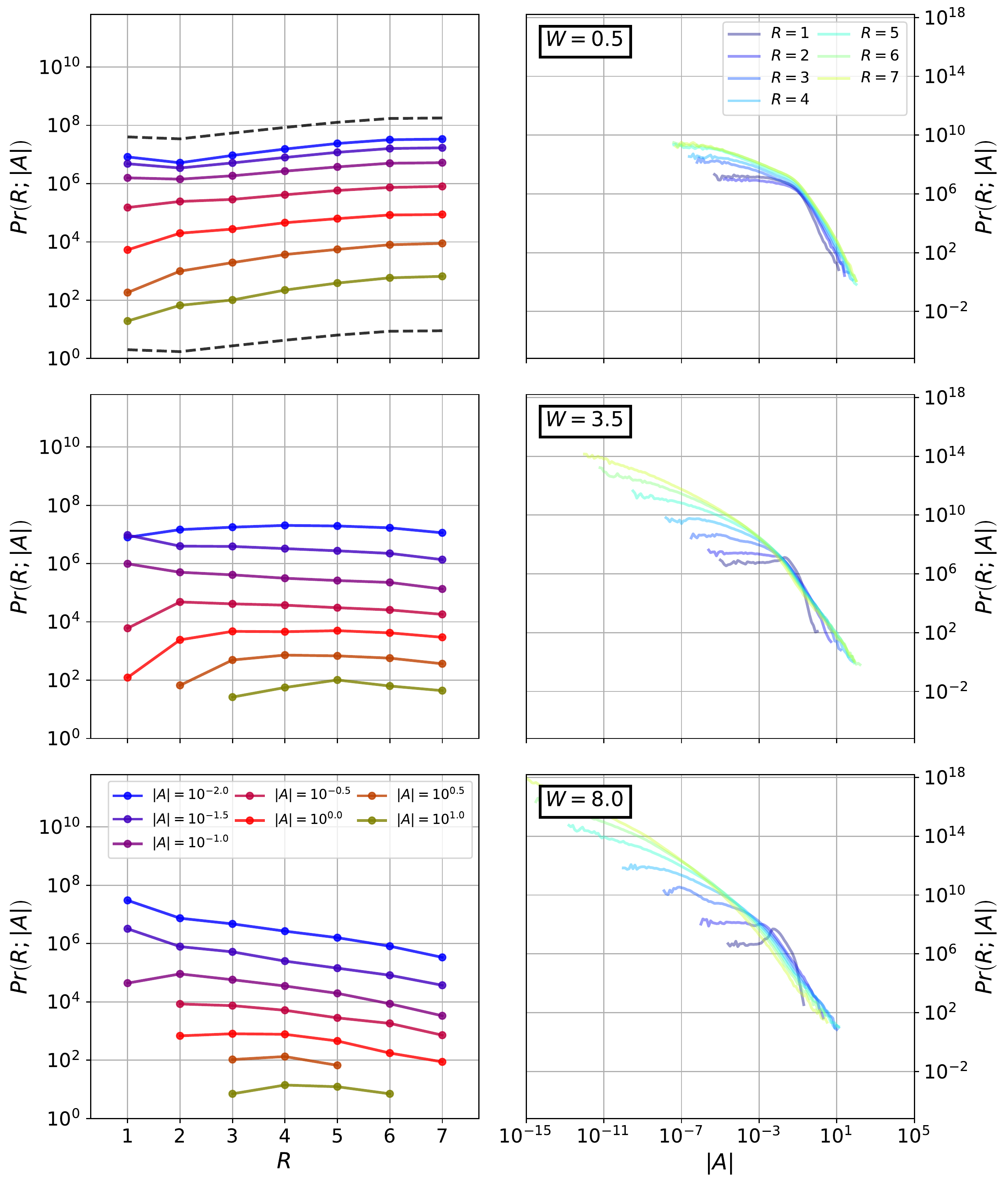}
\caption{\label{fig:invariance_L8} Slices of $Pr(R; |A|)$ for $L = 8$.
This is similar to Fig.\ref{fig:invariance} of the main text.} 
\end{figure}

\begin{figure}[t]
\includegraphics[width=1.0\columnwidth]{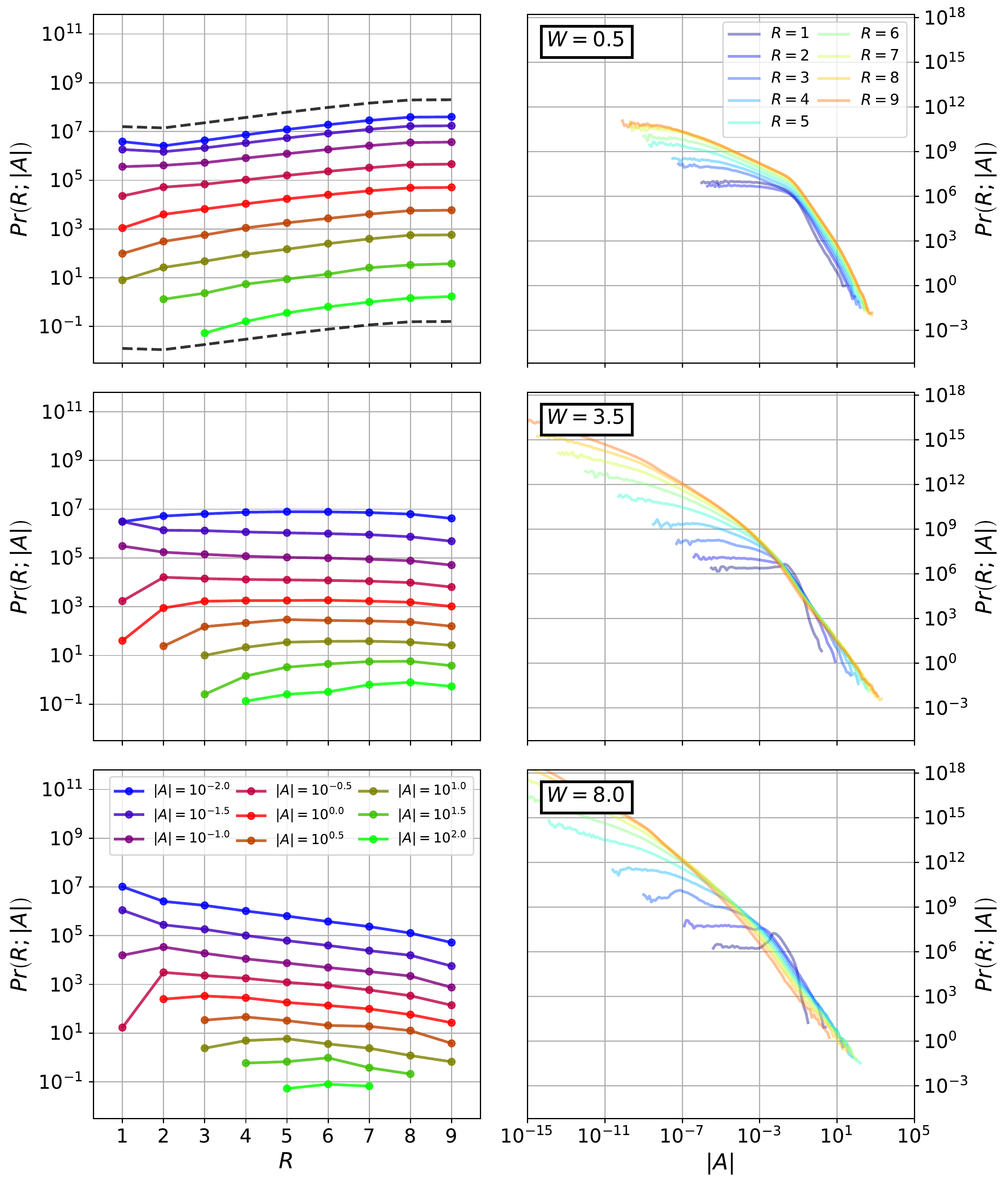}
\caption{\label{fig:invariance_L10} Same as Fig.~\ref{fig:invariance_L8} for $L = 10$.} 
\end{figure}

In Figs.~\ref{fig:invariance_L8} and \ref{fig:invariance_L10} we show the equivalent of Fig.~\ref{fig:invariance} of the main text for $L = 8, 10$.

\section{Alternative view of the collapse of $Pr(R; |A|)$ at large $|A|$ to a universal function at the transition}
\label{sec:collapse}

\begin{figure*}[t]
\includegraphics[width=2.0\columnwidth]{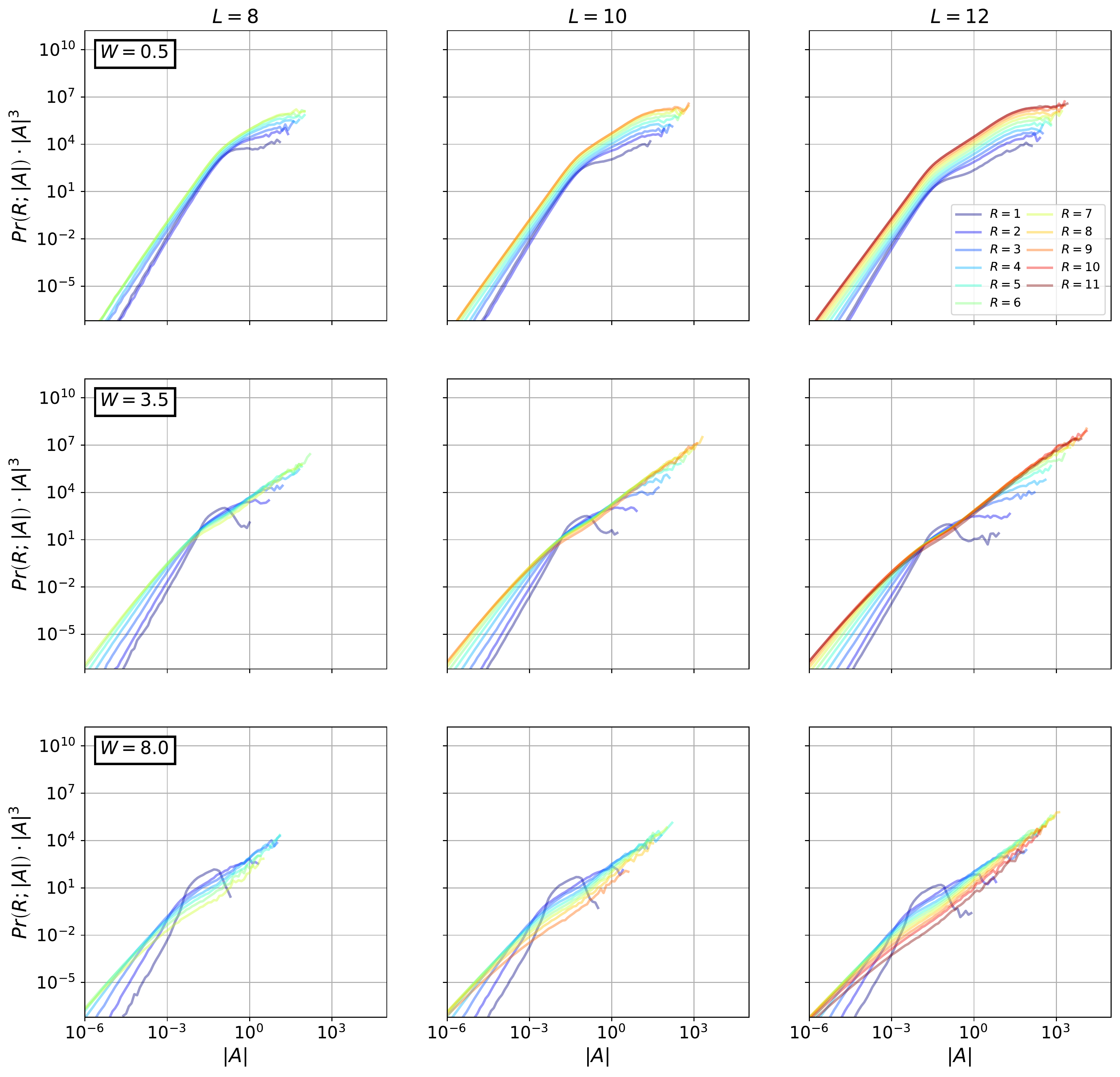}
\caption{\label{fig:hist_shifted} Slices of fixed $R$ of $Pr(R; |A|) |A|^3$ as a function of $|A|$ for different system sizes.
We can appreciate a collapse of the curves onto a universal functional form at the transition $W \approx 3.5$ for large enough $|A|$ and before each curve hits its $R$-dependent cutoff.
} 
\end{figure*}

The probability distribution $Pr(R; |A|)$ collapses to a universal function at large enough $|A|$ at the ergodic-MBL transition.
This is difficult to appreciate clearly from Fig.~\ref{fig:invariance} of the main text.
In Fig.~\ref{fig:hist_shifted} we plot slices of $Pr(R; |A|) |A|^3$ at fixed $R$, which simply shifts the first derivative of the curves (in a log-log scale) by 3.
This view allows us to appreciate the collapse at the transition (see $W = 3.5$ for $L = 12$) of the curves for different ranges onto a universal functional form at large enough $|A|$ and before they hit their range-dependent cutoff.
Note that the curves are modified equally for all ranges, and so a collapse of $Pr(R; |A|) |A|^3$ implies automatically a collapse of $Pr(R; |A|)$.

\section{Alternative visualization of $Pr(R; |A|)$}
\label{sec:alternative_pr}

\begin{figure*}[t]
\includegraphics[width=2.0\columnwidth]{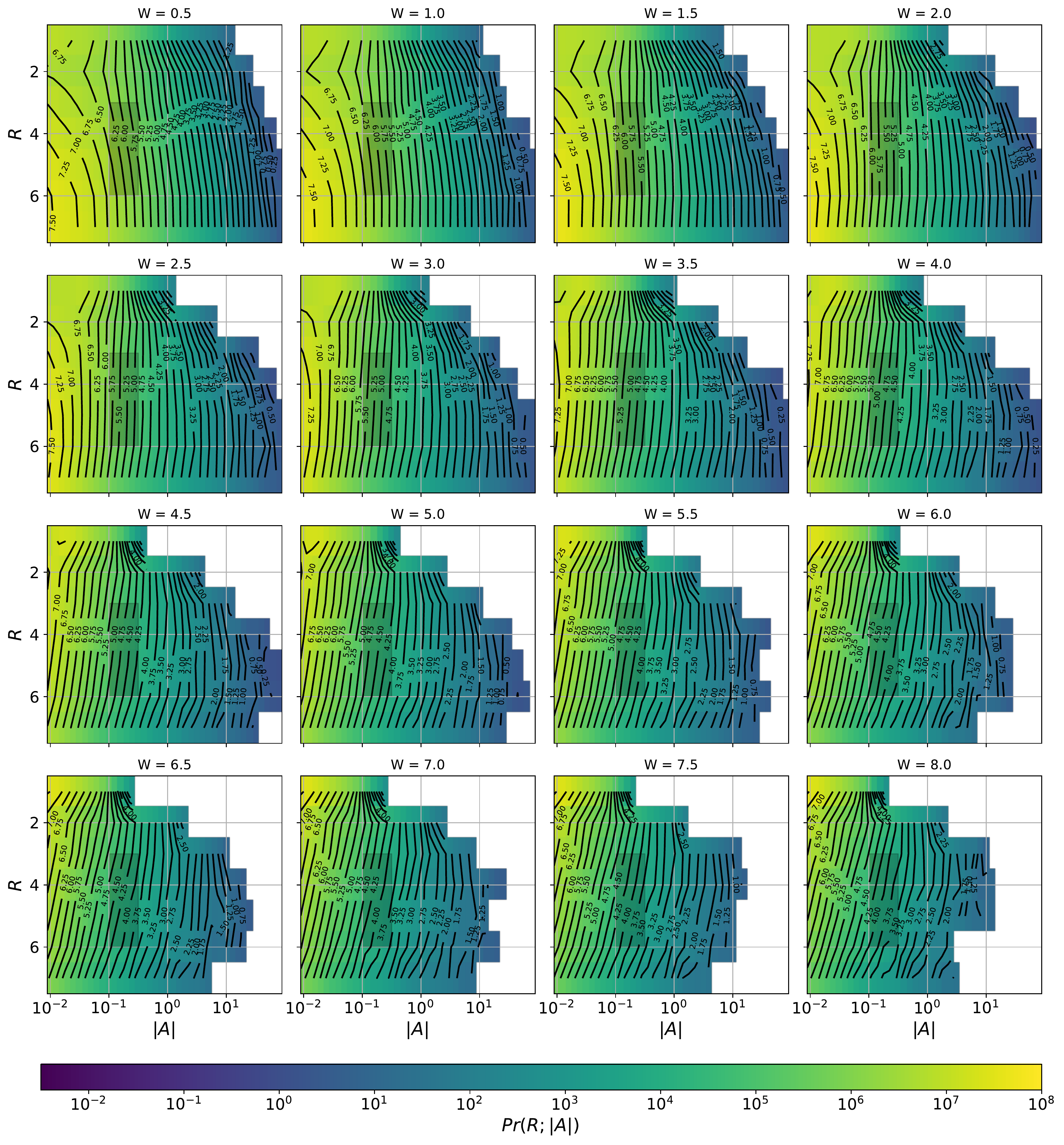}
\caption{\label{fig:pr_colormap_L8} Colormap of $Pr(R; |A|)$ with contour lines for $L = 8$.
For visibility in the plots, the values on the contour lines correspond to $\log_{10} [Pr(R; |A|)]$.
The exponent $\alpha$ was computed from exponential fits to slices of $Pr(R; |A|)$ on the shaded area.} 
\end{figure*}

\begin{figure*}[t]
\includegraphics[width=2.0\columnwidth]{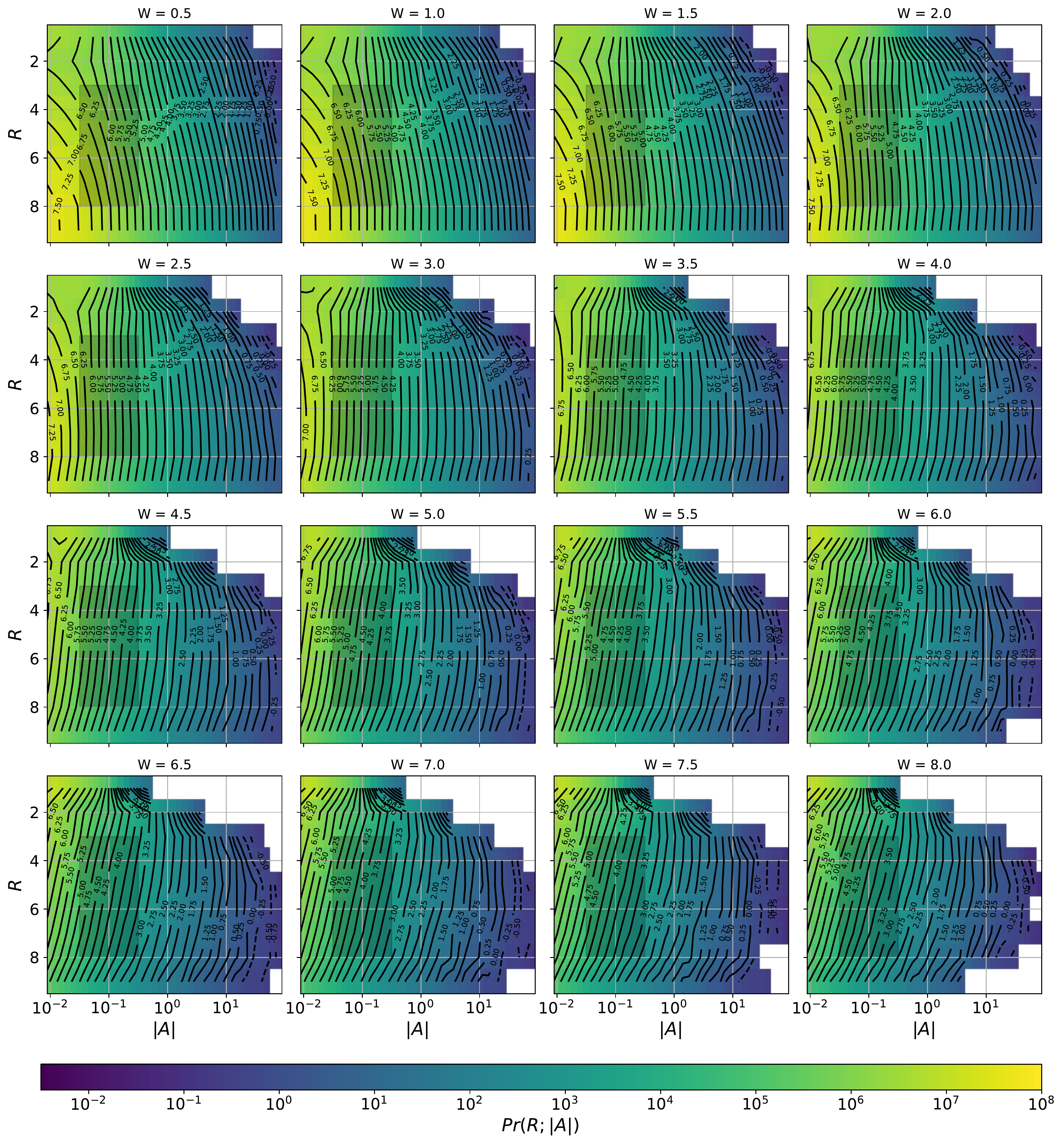}
\caption{\label{fig:pr_colormap_L10} Same as Fig.~\ref{fig:pr_colormap_L8} for $L = 10$.} 
\end{figure*}

\begin{figure*}[t]
\includegraphics[width=2.0\columnwidth]{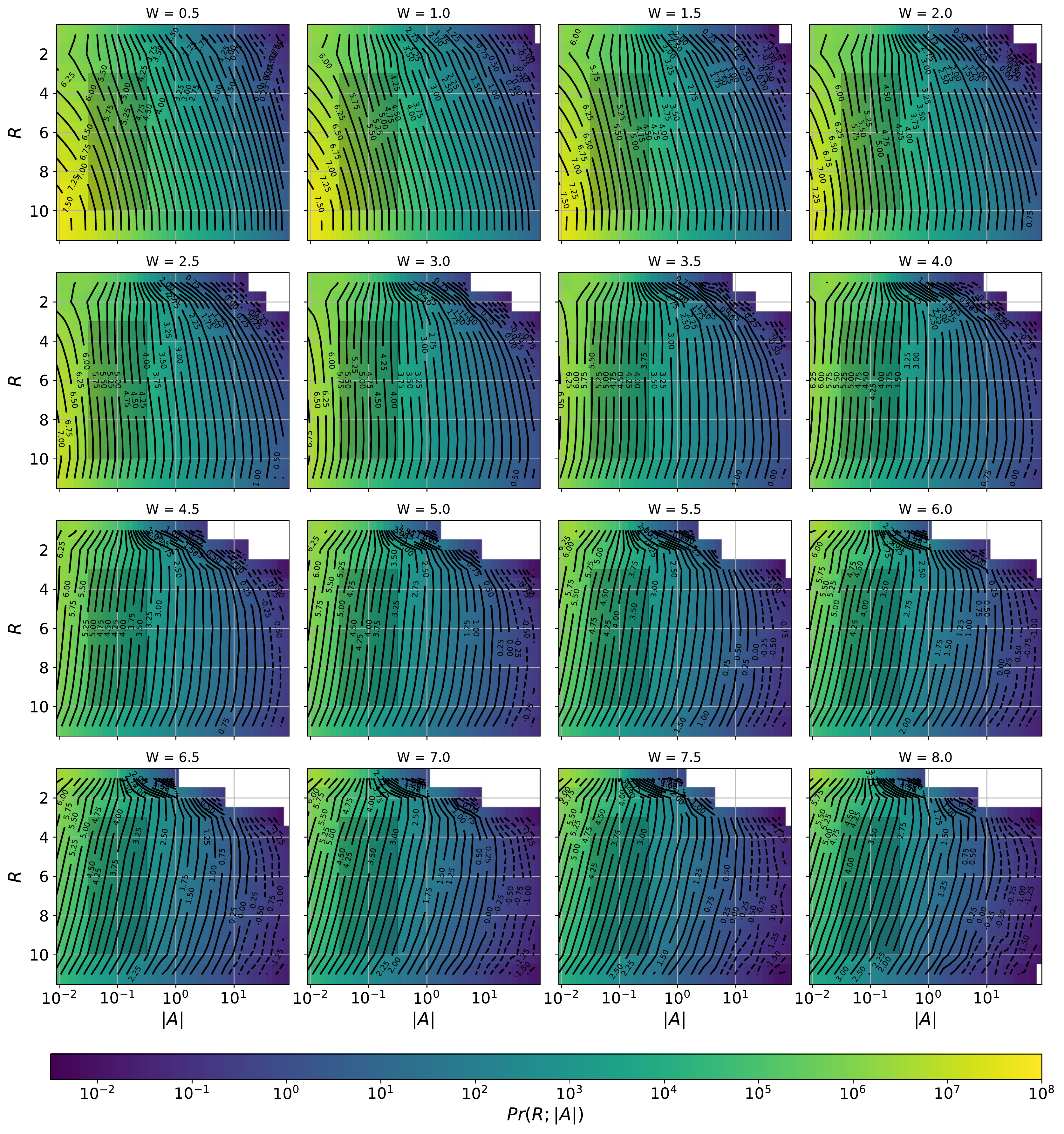}
\caption{\label{fig:pr_colormap_L12} Same as Fig.~\ref{fig:pr_colormap_L8} for $L = 12$.} 
\end{figure*}

Here we present an alternative visualization of the joint probability distribution $Pr(R; |A|)$.
Instead of showing slices at constant $R$ or $|A|$, we show a colormap of the distribution with contour lines (Figs.~\ref{fig:pr_colormap_L8}, \ref{fig:pr_colormap_L10}, and \ref{fig:pr_colormap_L12}).
We can see that at large $|A|$ the probability that a pair of eigenstates hybridize over a particular range becomes range invariant at the transition, with contour lines becoming vertical (see the case of $L = 12$, $W = 3.5$, over the shaded area).
We can also appreciate finite size effects at the largest $R$ for each $L$, as well as $R$-dependent cutoffs for the smallest ranges, where $Pr(R; |A|)$ drops its universal behavior.
Finally, equally spaced contour lines imply a power law in $Pr(R; |A|) \propto |A|^beta$) at constant $R$, as discussed in the main text.

The values of $\alpha$ presented in the main text were obtained from fits ($\log[Pr(R; |A|)] = \alpha R + const.$) over slices at constant $|A|$ over the shaded areas, \emph{i.e.}, avoiding finite size effects and constraining the fits to large $|A|$, while avoiding $R$-dependent cutoffs.

\section{Combinatorial counting of clusters as a function of range}
\label{sec:combinatorial}

Given two eigenstates and their respective sets of pseudo-spins, they can disagree over a subset or clusterthat spans a region of range $R$.
Here we derive an analytical formula for the number of possible clusters for a fixed $R$, $N(R)$.

First, note that two sets of quantum numbers have to disagree in an even number of sites, given the fact that we are working on the zero magnetization sector.
Furthermore, the magnetization of the cluster of disagreeing quantum numbers must also be zero, \emph{i.e.}, have an equal number up and down pseudo-spins.
We can now proceed with the counting.

For a fixed range $R$, clusters can take any order $o = 2, 4, 6, \ldots R$ (number of pseudo-sites).
For each order, and ignoring cyclic permutations of the bit-strings, two of the disagreeing pseudo-spins have to be ``pinned'' at the left and right ends of the range $R$ interval, leaving only freedom to the inner $o - 2$ pseudo-spins to be placed at different intermediate sites; there are therefore $\binom{R - 2}{o - 2}$ ways to place them.
This yields:
\begin{align}
    \label{eq:places}
    \sum_{\text{even } o}^R \binom{R -2}{o - 2}
\end{align}
ways to place disagreeing pseudo-spins while forming a range $R$ cluster.
There is another factor to take into account: once the $o$ disagreeing pseudo-spins are placed, their orientations have to be chosen.  Given that the magnetization of the $o$ pseudo-spins is zero, there are $\binom{o}{o / 2}$ ways to arrange them, times the number of ways the agreeing pseudo-spins can be arranged, which is $\binom{L - o}{(L - o) / 2}$, and where $L$ is the total number of sites in the system.
We then get that the scaling is:
\begin{align}
    \label{eq:incomplete_combinatorial_scaling}
    \propto \sum_{\text{even } o}^R \binom{R -2}{o - 2} \binom{o}{o / 2} \binom{L - o}{(L - o) / 2} \text{.}
\end{align}

Finally, we have to account for tranlations of the bitstrings on the chain. Given open boundary conditions, the number of cyclic permutations allowed for a particular cluster of range $R$ is equal to $L - 1 - R$. We finally have the right scaling:
\begin{align}
    \label{eq:combinatorial_scaling}
    N_L(R) = \sum_{\text{even } o}^R &\binom{R -2}{o - 2} \binom{o}{o / 2}  \nonumber \\
    &\binom{L - o}{(L - o) / 2} \left( L - 1 - R \right) \text{.}
\end{align}
In the thermodynamic limit, this goes as $\propto 2^R$.
The curve $N_L(R)$ was plotted in Fig.~\ref{fig:invariance} (top-left, dashed lines) for convenient comparison with $Pr(R; |A|)$ as a function of $R$ deep in the ergodic phase.
While the value of $\alpha$ (see main text) is $\log(2)$ in the thermodynamic limit, for finite size systems it can be computed numerically from Eq.~\eqref{eq:combinatorial_scaling}.

\section{Relation of $\xi$ to other correlation lengths}
\label{sec:other_xis}

While there are a number of other correlation lengths in the literature which scale as $\log(W)$, not all of them exhibit $B=1/2$; note that $\log(W)$ is the scaling expected for the correlation length of an Anderson insulator at strong disorder.
Since rescaling our units of length multiplicatively by $s$ rescales our inverse correlation length as $1 / s$, it is a reasonable conjecture that we could fit other correlation lengths as $\xi^{-1}(W) = -B/s \log(4W/W_c)$.
Using this ansatz, we test our results against two other correlation lengths.
These are not chosen generically but are selected to be ones which both show the expected logarithmic scaling and have small errors.
First, we consider Ref.~\cite{varma_length_2019} (IOM ($xX$) from their Fig.~1); this correlation length scales logarithmically but with $B/s\approx 1$;  fitting their data we get a $W_c$ of 4.68.
Finally, we consider the freezing exponent of the recent preprint, Ref.~\cite{laflorencie_chain_2020}, which is inversely proportional to the correlation length and yields a critical value of $W_c \approx 4.44$ with a $B/s\approx 1.67$.
This is consistent with an independent estimate of the $W_c$ in their paper.

\section{All $\beta$ fits}
\label{sec:beta_fits}

\begin{figure*}[t]
\includegraphics[width=1.4\columnwidth]{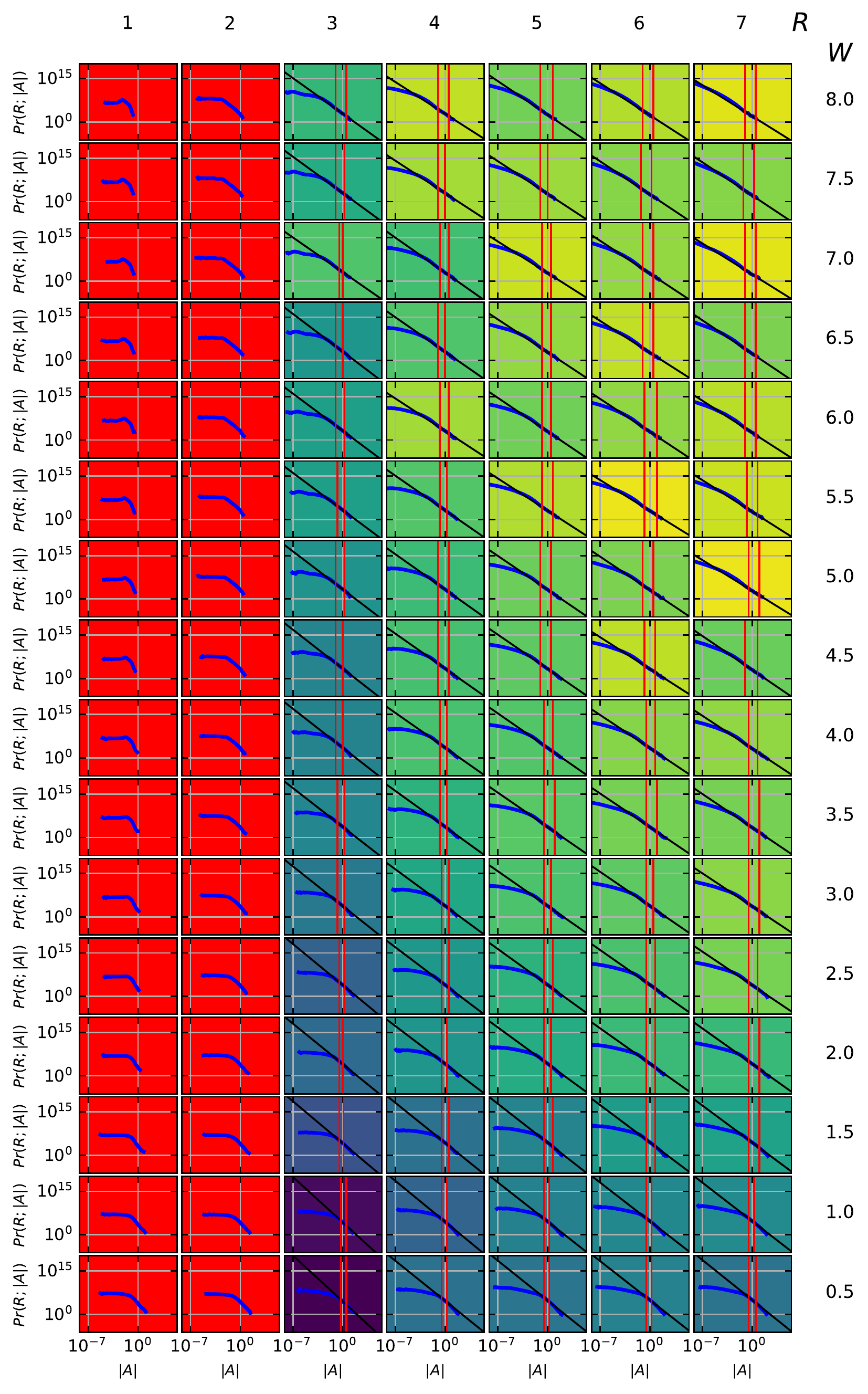}
\caption{\label{fig:beta_fits_L8} Empirical fits of $Pr(R; |A|)$ to a power law $|A|^\beta$ for systems of size $L = 8$.
The exponent $\beta$ was reported in Fig.~\ref{fig:beta} the main paper.
Due to the noisy nature of the data, we make a linear fit on the log-log plot on a region chosen ad hoc for each case (see red vertical lines).} 
\end{figure*}

\begin{figure*}[t]
\includegraphics[width=1.7\columnwidth]{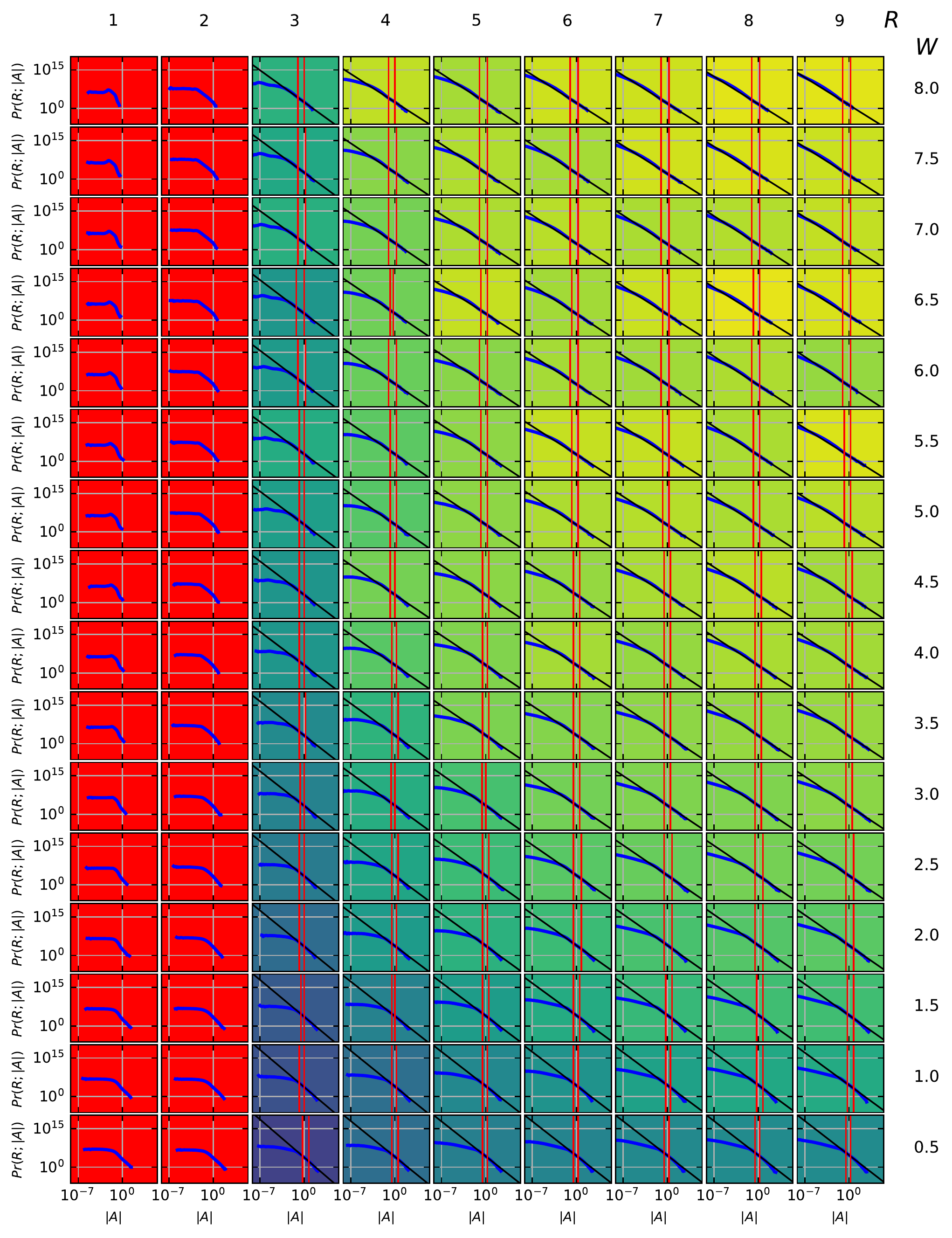}
\caption{\label{fig:beta_fits_L10} Same as Fig.~\ref{fig:beta_fits_L10} for $L = 10$.} 
\end{figure*}

\begin{figure*}[t]
\includegraphics[width=2.0\columnwidth]{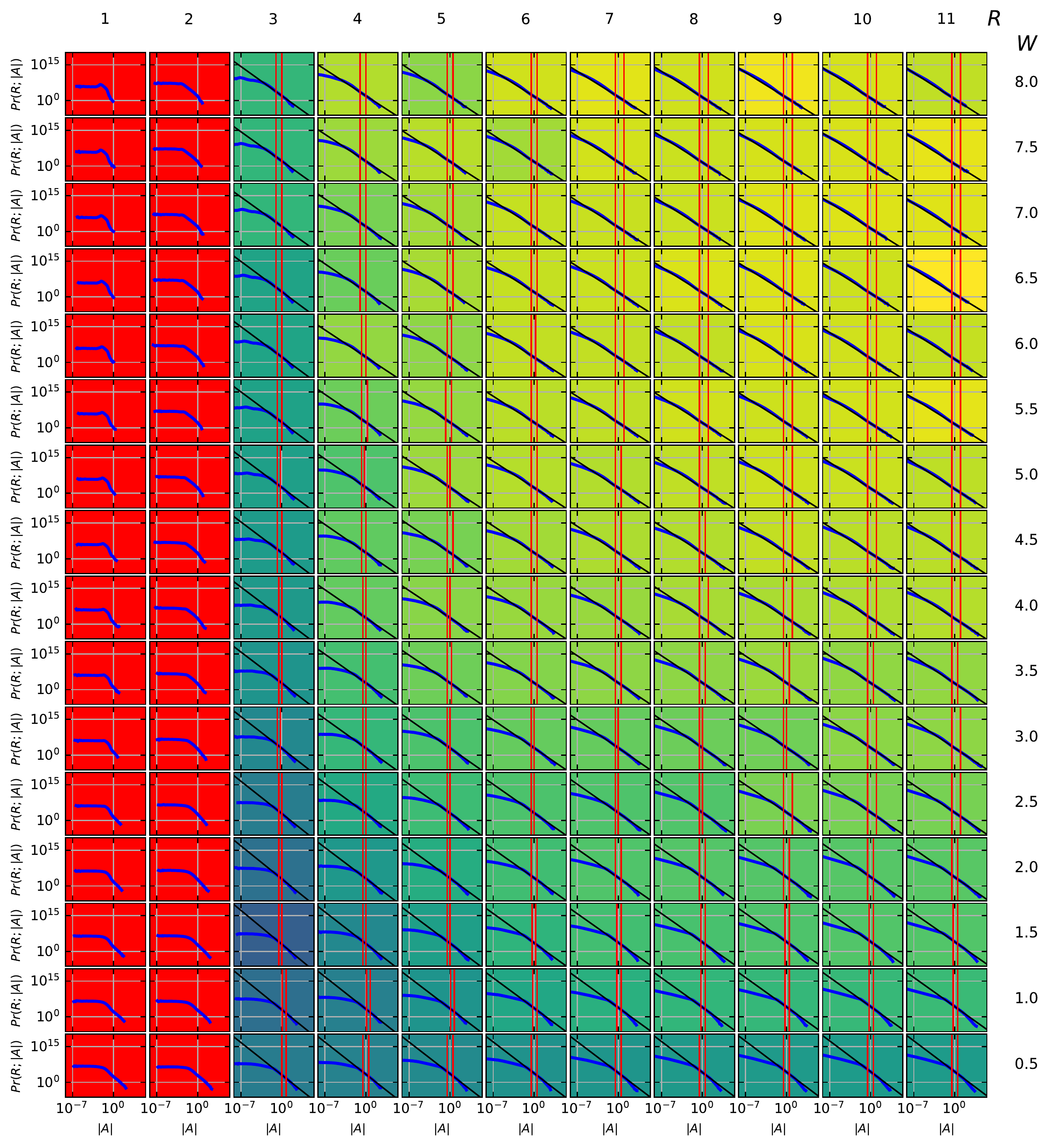}
\caption{\label{fig:beta_fits_L12} Same as Fig.~\ref{fig:beta_fits_L12} for $L = 12$.} 
\end{figure*}

Here we provide all linear fits reported in Fig.~\ref{fig:beta} of the main text for the computation of the exponent $\beta$.
Fig.~\ref{fig:beta_fits_L8} reports the all fits for $L = 8$, Fig.~\ref{fig:beta_fits_L10} reports $L = 10$ fits, and Fig.~\ref{fig:beta_fits_L12} reports $L = 12$ fits.
We extract $\beta$ for a linear fit in the log-log plot of $Pr(R; |A|)$ as a function of $|A|$ for fixed $R$.
In order to decrease noise in the fit, we perform fits over several intervals around a central interval in $|A|$ (shown by the two red vertical lines); we then average the results. 
In all cases, the background of the plots follows the same colormap as in Fig.~\ref{fig:beta} of the main text.
Ranges $R = 1, 2$ (red panels) are difficult to interpret, since the region typically fitted blends with the cutoff final region. We choose to ignore these.

\clearpage


\end{document}